\documentclass[%
 aip,
 amsmath,amssymb,
 reprint,%
 citeautoscript, %
]{revtex4-1}

\usepackage{graphicx}
\usepackage{dcolumn}
\usepackage{bm}

\newcommand{\ket}[1]{\left|#1\right\rangle}
\newcommand{\braket}[1]{\left|#1\right\rangle\left\langle #1 \right|}
\newcommand{\dbraket}[2]{\left|#1\right\rangle\left\langle #2 \right|}
\newcommand{\trace}[1]{\text{Tr} \left\{ #1 \right\} }
\newcommand{\traceS}[1]{\text{Tr}_S \left\{ #1 \right\} }
\newcommand{\traceB}[1]{\text{Tr}_B \left\{ #1 \right\} }

\begin{document}
\raggedbottom
\preprint{AIP/123-QED}

\title{Vibronic coupling in energy transfer dynamics and two-dimensional electronic-vibrational spectra}

\author{Eric A. Arsenault}
\altaffiliation{These authors contributed equally.}
\affiliation{\looseness=-1Department of Chemistry, University of California, Berkeley, CA 94720, USA}
\affiliation{\looseness=-1Kavli Energy Nanoscience Institute at Berkeley, Berkeley, CA 94720, USA}
\affiliation{\looseness=-1Molecular Biophysics and Integrated Bioimaging Division, Lawrence Berkeley National Laboratory, Berkeley, CA 94720, USA}
 
\author{Addison J. Schile}
\altaffiliation{These authors contributed equally.}
\affiliation{\looseness=-1Department of Chemistry, University of California, Berkeley, CA 94720, USA}
\affiliation{\looseness=-1Chemical Sciences Division, Lawrence Berkeley National Laboratory, Berkeley, CA 94720, USA}

\author{David T. Limmer}
  \email{dlimmer@berkeley.edu}
\affiliation{\looseness=-1Department of Chemistry, University of California, Berkeley, CA 94720, USA}
\affiliation{\looseness=-1Kavli Energy Nanoscience Institute at Berkeley, Berkeley, CA 94720, USA}
\affiliation{\looseness=-1Chemical Sciences Division, Lawrence Berkeley National Laboratory, Berkeley, CA 94720, USA}
\affiliation{\looseness=-1Materials Science Division, Lawrence Berkeley National Laboratory, Berkeley, CA 94720, USA}

\author{Graham R. Fleming}
  \email{grfleming@lbl.gov}
\affiliation{\looseness=-1Department of Chemistry, University of California, Berkeley, CA 94720, USA}
\affiliation{\looseness=-1Kavli Energy Nanoscience Institute at Berkeley, Berkeley, CA 94720, USA}
\affiliation{\looseness=-1Molecular Biophysics and Integrated Bioimaging Division, Lawrence Berkeley National Laboratory, Berkeley, CA 94720, USA}

\date{\today}

\begin{abstract}

We introduce a heterodimer model in which multiple mechanisms of vibronic coupling and their impact on energy transfer can be explicitly studied. We consider vibronic coupling that arises through either Franck-Condon activity in which each site in the heterodimer has a local electron-phonon coupling and as Herzberg-Teller activity in which the transition dipole moment coupling the sites has an explicit vibrational mode-dependence. We have computed two-dimensional electronic-vibrational (2DEV) spectra for this model while varying the magnitude of these two effects and find that 2DEV spectra contain static and dynamic signatures of both types of vibronic coupling.
Franck-Condon activity emerges through a change in the observed excitonic structure while Herzberg-Teller activity is evident in the appearance of significant side-band transitions that mimic the lower-energy excitonic structure. A comparison of quantum beating patterns obtained from analysis of the simulated 2DEV spectra shows that this technique can report on the mechanism of energy transfer, elucidating a means of experimentally determining the role of specific vibronic coupling mechanisms in such processes.

\end{abstract}

\maketitle

\section{\label{sec:intro}Introduction}

Elucidating the mechanisms of quantum mechanical energy transfer has fundamental implications for the way we understand natural light-harvesting and develop artificial analogs.\cite{scholes2011lessons}
Previous experimental studies on natural systems\cite{engel2007evidence,calhoun2009quantum,myers2010two} have been unable, however, to clearly establish the  mechanism of energy transfer that leads to quantum efficiencies approaching unity\cite{blankenship2014molecular} and have launched long-standing debates obfuscating the role of observed electronically and/or vibrationally coherent phenomena in the transfer process.\cite{abramavicius2010quantum, panitchayangkoon2011direct,tempelaar2014vibrational,rolczynski2018correlated,plenio2013origin,butkus2014vibronic,christensson2012origin,polyutov2012exciton,jonas2018vibrational,cina2004vibrational,romero2014quantum,fuller2014vibronic, arsenault2020vibronic,higgins2021photosynthesis,novoderezhkin2017exciton,dean2016vibronic,thyrhaug2018identification,ma2018vibronic} It has been postulated that these coherent processes may not actually serve any purpose in the overall energy transfer mechanism.\cite{fujihashi2015impact,cao2020quantum} This ambiguity largely surrounds the lack of consistent treatment of electronic-vibrational coupling in energy transfer models, which we address through a simplified heterodimer model in this paper. It has been shown that explicit details of the vibronic coupling mechanism can have a large influence on the overall dynamics.\cite{yeh2019elucidation, zhang2016effects, duan2019ultrafast, duan2020intramolecular} Also contributing to the uncertainty is that the distinguishing features between vibronic mixing mechanisms in coupled systems can be subtle in electronic spectroscopies\cite{seibt2018treatment, zhang2016effects,prall2005anti}---and are only further obscured in the complex, congested spectra of experimental realizations. 

Recently, two-dimensional electronic-vibrational (2DEV) spectroscopy has emerged as a candidate experimental technique that can directly observe the correlated motion of electronic and nuclear degrees-of-freedom and their role in energy transfer.\cite{oliver2014correlating} Indeed, initial studies on photosynthetic complexes, such as light-harvesting complex II (LHCII), showed promise in utilizing this technique to unravel the dynamics of energy transfer between different chromophores owing to the improved spectral resolution and structural details afforded via probing vibrational modes.\cite{lewis2016observation} Subsequent 2DEV measurements have shown evidence of vibronic mixing in and its facilitation of ultrafast energy transfer in LHCII.\cite{arsenault2020vibronic} 
In the latter, the 2DEV spectra showed rich vibrational structure corresponding to the dominant electronic excitations which exhibited oscillatory dynamics reminiscent of non-Condon effects found in previous transient absorption measurements.\cite{prall2005anti,yoneda2019non,kano2002observation} 
These oscillations were also found to be present at slightly higher-energy excitations to vibronically mixed states. 
In this case, the clear similarity in the quantum beating patterns between these higher-lying states and the dominant, more electronically mixed excitations, was speculated to be indicative of rapid energy relaxation due to vibronic mixing. Here we develop a strategy to simulate these general effects in 2DEV spectra and connect them to vibronic coupling mechanisms of energy transfer. Further 2DEV studies on LHCII, involving excitation well-beyond the dominant absorption bands, showed the same rapid energy relaxation, but with a significant polarization-dependence.\cite{arsenault2020role} 
With polarization control, the dynamics of vibronic excitations, exhibiting much more rapid energy transfer, were disentangled from purely electronic excitations with significantly slower energy transfer. 
Not only does this polarization-dependence isolate the role of vibronic mixing on the rate of energy transfer, it potentially rules out the role the protein environment has on enhancing rapid energy transfer and suggests a predominant contribution from \textit{intra}molecular modes to the underlying energy transfer mechanism. 

To date, theoretical work regarding the 2DEV signals of coupled systems, while informative, has been restricted to systems that have a only have a single vibrational mode per monomeric unit.\cite{lewis2015method,wu2019two,bhattacharyya2019two} An interpretation of the origin of the vibronic coupling observed in these recent findings is, therefore, lacking. Particularly, the relative infancy of 2DEV spectroscopy makes assigning vibronic mixing to direct electron-nuclear coupling or non-Condon effects in the experiments difficult as this requires the development of multimode models. In this paper, we bridge this gap between vibronic coupling mechanisms and analysis of the experimental measurements by directly simulating the 2DEV spectra of a minimal model vibronically coupled heterodimer while controlling various vibronic coupling mechanisms. By utilizing a model system, we are able to isolate the role that different vibronic coupling mechanisms have on the structure of the excitonic states that are electronically excited in typical experiments and show how that structure is identifiable in 2DEV spectroscopy both statically and dynamically. We further compare these signatures to the population dynamics, which demonstrates the ability to directly link the mechanism of energy transfer with spectral observables and connects model systems to potential \textit{ab initio} simulations for which only simple observables like the populations are available. 

The remainder of this paper is organized as follows. In Section \ref{sec:theory}, we introduce a model vibronic heterodimer and the formalism we use for computing linear absorption and 2DEV spectra. We analyze the static and dynamical signatures of vibronic coupling in the spectra in Sections \ref{sec:static} and \ref{sec:dynamics}, respectively. Concluding remarks and directions for future work are provided in Section \ref{sec:conclusion}. 

\section{\label{sec:theory}Theory}

In this section, we introduce a minimal vibronically coupled heterodimer model and the theoretical formalism by which we simulate spectra. We utilize an open quantum system approach to describe the heterodimer in contact with a thermal bath given by the total Hamiltonian, $H = H_S + H_B + H_{SB}$, where $H_S$ is the system Hamiltonian of the heterodimer, $H_B$ is the bath Hamiltonian, and $H_{SB}$ is the system-bath Hamiltonian describing their interactions. This approach offers an exact description of the most strongly-coupled system degrees-of-freedom with a simple treatment of relevant environmental effects that induce dissipation and dephasing in the system. 

\subsection{\label{sec:model}Model Hamiltonian}

\begin{figure*}
\begin{center}
\includegraphics[scale=0.5238]{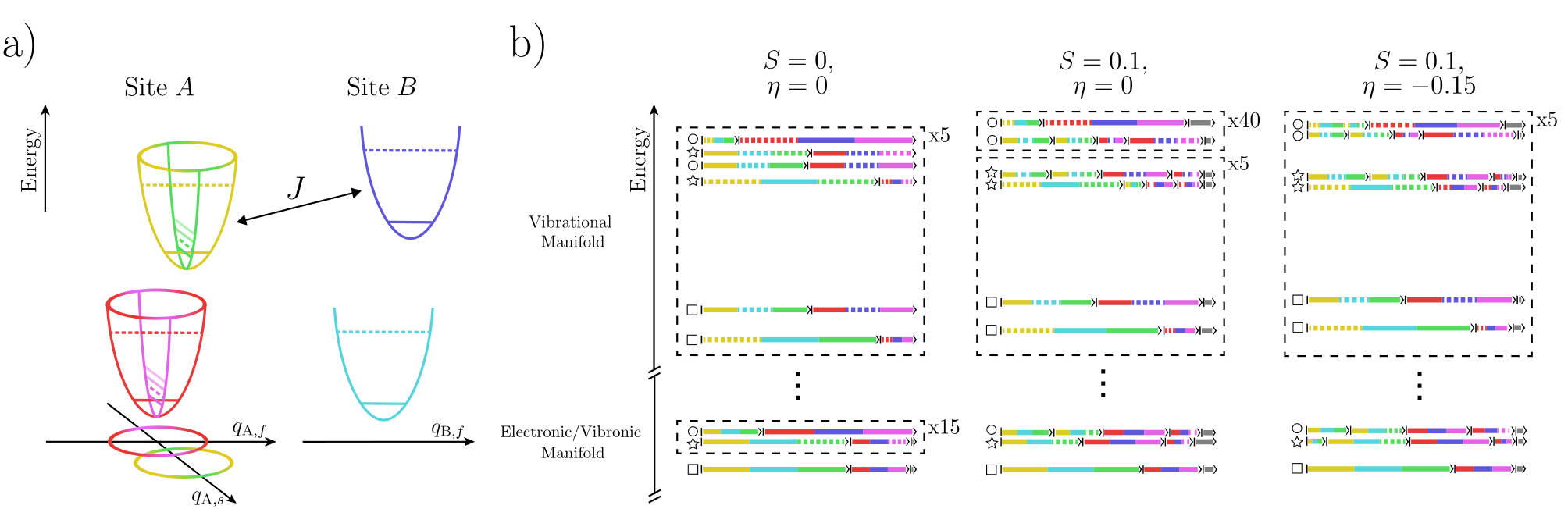}
\end{center}
\caption{\label{fig:fig1} a) Schematic of the model consisting of two sites, $A$ and $B$, each with one electronic degree-of-freedom and one high-frequency vibrational mode. For site $A$ ($B$), the high-frequency mode is shown in red (light blue) in the ground state and yellow (dark blue) in the excited state. Horizontal lines in each harmonic potential indicate vibrational levels where dashed lines, specifically, indicate one vibrational quantum. Site $A$ includes an additional low-frequency mode, shown in pink in the ground state and green in the excited state with line-markings corresponding to different vibrational excited states. b) Simplified eigenenergy level diagram arising from electronic coupling, $J$, between the excited state manifolds of sites $A$ and $B$ for each of the three models considered here. For simplicity, the ground state manifold has been omitted and only the three lowest excitonic states in the excited electronic/vibronic manifold as well as the corresponding vibrational levels have been illustrated. The relative site contributions for these levels are also shown by the length of each ket, $``\lvert \; \rangle"$, with site-specific color-coding following a). Dashed lines indicate a vibrational excitation. Site contributions of $<5\%$ are grouped together and denoted in gray. Shapes on the left hand side of the energy levels in the electronic/vibrational denote the main excited state absorption (ESA) transitions while shapes in the vibrational manifold denote the states to which the main ESA excitations are excited by vibrational pulses.}
\end{figure*}

The system (depicted in Fig. \ref{fig:fig1}a) is comprised of two chromophores (herein referred to as sites $A$ and $B$) each consisting of a local ground and excited electronic state and local intramolecular modes. These chromophores, in the context of natural light-harvesting, could be considered distinct pigments in a protein or two of the same pigments with different protein binding properties that statically change the characteristics of the local Hamiltonians. We restrict the system Hamiltonian to the ground state ($G$) and singly-excited state manifold, thus containing three electronic states of the form,
\begin{align}
H_S &= (h_A^g + h_B^g) \braket{G} + (h_A^e + h_B^g + \epsilon) \braket{A} \nonumber \\
&+ (h_A^g + h_B^e + \Delta E + \epsilon) \braket{B} + J ( \dbraket{A}{B} + \text{h.c.} ),
\end{align}
where we have implied the Kronecker product structure of the $A$ and $B$ local Hamiltonians applying on their local vibrational subspace. The electronic state $\ket{A}$ ($\ket{B}$) refers to the state when site $A$ ($B$) is excited and site $B$ ($A$) is in its ground state. Here, the ground state is uncoupled to and energetically separated from the excited states by an excitation energy, $\epsilon$, which may be removed without loss of generality. The excited states comprise a two-level system in the electronic subspace that has an energy difference denoted by $\Delta E$ and an electronic coupling denoted by $J$. In this two-level subsystem it is useful to consider the excitonic gap, which is equivalent to a Rabi frequency given by $\hbar \Omega_R = \sqrt{\Delta E^2 + 4 J^2}$ that determines the timescale of electronic oscillations between the excited states. 

Each site has a ground ($g$) and excited ($e$) state where the local Hamiltonians acting on the site vibrational subspaces have the form
\begin{align}
h_I^i &= \frac{\hbar \omega_{I,i,f}}{2} \left(p_{I,f}^2 + (q_{I,f}-\delta_{ie}\sqrt{2S_f})^2\right) \nonumber \\
&+ \delta_{IA} \frac{\hbar \omega_{I,s}}{2} \left(p_{I,s}^2 + (q_{I,s}-\delta_{ie}\sqrt{2S})^2\right)
\end{align}
where $I=A,B$ denotes the chromophore site, $i=g,e$ denotes the electronic state of the site,  $\delta_{ij}$ denotes the Kronecker delta, and the $q$ and $p$ are the position and momentum operators, respectively of high-frequency, $f$, and low-frequency, $s$, modes. Each site contains one high-frequency ($ \omega_{I,i,f}/\Omega_R \gg 1$) local intramolecular modes with a distinct site- and electronic-state-dependent frequency. These high-frequency modes are slightly displaced in the excited states and thus have a small, but non-zero Huang-Rhys factor, $S_f$, which we will consider fixed throughout this study. Vertical excitations and electronic transitions are, however, still dominated by transitions that leave the vibrational states of these modes unchanged.  Coupled to site $A$ only is also a low-frequency mode that is nearly-degenerate with the excitonic gap, $ \omega_{A,s} \approx \Omega_R$. In practice, this mode could be considered an intramolecular mode with significant local site electron-phonon coupling. This mode is also shifted in the excited state of the $A$ chromophore with a non-zero Huang-Rhys factor, $S$, however due to the resonance with the excitonic gap, this displacement induces significant vibronic mixing by coupling different vibrational states in vertical excitations from the ground state or electronic transitions between the $A$ and $B$ sites. Thus, $S$ can be varied to tune the strength of the vibronic coupling mechanism through what we herein refer to as Franck-Condon (FC) activity. We note that in this work, two vibrational levels per high-frequency mode and four vibrational levels for the low-frequency mode were required for convergence. Additionally, we have restricted the model to the ground- and singly- excited vibrational state manifold with respect to the subspace of the high-frequency modes for a total system Hilbert space dimension of 36.

The electronic coupling is considered to arise from a dipole-dipole interaction between the excited states of the two chromophores,
\begin{align}
J = \frac{\kappa}{r^3} \mu_A \mu_B,
\end{align}
where $\mu_{A(B)}$ is the magnitude of the transition dipole moment (TDM) for the $A$ ($B$) site, $r$ is the distance between the two chromophores, and $\kappa$ is a factor accounting for the relative orientation of the chromophores. We assume here that the distance, relative orientation, and TDM of the $B$ chromophore are fixed ($r=r_0$, $\kappa=\kappa_0$, and $\mu_B = \mu_{B0}$, respectively), while the TDM of the $A$ chromophore depends linearly on the low-frequency mode,
\begin{align}
\mu_A (q_{A,s}) = \mu_{A0} \left( 1 + \sqrt{2} \eta q_{A,s} \right),
\end{align}
where $\mu_{A0}$ is the static contribution to the dipole moment. The mode-dependence arises as a non-Condon effect, that is,
\begin{align}
\sqrt{2} \mu_{A0} \eta = \left( \frac{\partial \mu_A}{\partial q_{A,s}} \right),
\end{align}
where $\eta$ is a dimensionless parameter controlling the strength of this effect. We note that because the electronic states have the same symmetry there is no strict symmetry requirement here for the HT active mode. \cite{albrecht1960forbidden} Under this assumption, the electronic coupling obtains the form
\begin{align}
J (q_{A,s}) = J_0 \left( 1 + \sqrt{2} \eta q_{A,s} \right),
\end{align}
where $J_0$ is the electronic coupling arising from the static contributions of the TDM at a fixed distance and orientation, $J_0 = \kappa_0 \mu_{A0} \mu_{B0} / r_0^3$,
and the non-Condon effect is given by $\sqrt{2} J_0 \eta q_{A,s}$. We consider here a system in the electronically coherent regime ($\Delta E = J_0$), which is typical for energy transfer dynamics in these chromophoric systems. Since $\eta$ is a dimensionless parameter and it enters directly in the TDM, it can be varied to systematically study Herzberg-Teller (HT) activity in this system.

The chromophoric system here is assumed to be weakly coupled to a set of environmental modes that describe the short- and long-range fluctuations of the environment. In particular we consider two sets of baths, an electronic set and a vibrational set, which are assumed to be independent due to disparity of the frequency of modes that couple to the separate electronic or vibrational degrees-of-freedom. The electronic baths independently couple to the electronically excited states through a dipolar coupling 
\begin{align}\label{eq:elbath}
\left( H_B + H_{SB} \right)_{\text{el}} = \sum_{I,n} \frac{\hbar \omega^{\text{el}}_{I,n}}{2} &\left[ \left( p^{\text{el}}_{I,n} \right)^2 \right. \nonumber \\
&+ \left. \left( q^{\text{el}}_{I,n} -  \frac{g^{\text{el}}_{I,n}}{\sqrt{2}} V_{I} \right)^2 \right],
\end{align}
where $V_I$ ($I=A,B$) are the dimensionless system dipole operators
\begin{align}
V_{I} = \left( 1 + \delta_{IA} \sqrt{2} \eta q_{I,s} \right) \braket{I} 
\end{align}
and the vibrational baths independently couple to the nuclear modes of the system
\begin{align}\label{eq:vibbath}
( H_B + H_{SB} &)_{\text{vib}} = \sum_{I,n} \left\{ \frac{\hbar \omega^{f}_{I,n}}{2} \left[ \left( p^{f}_{I,n}\right)^2 \left( q^{f}_{I,n} - g^{f}_{I,n} q_{I,f} \right)^2 \right] \right. \nonumber \\
&+ \left. \delta_{IA} \frac{\hbar \omega^{s}_{I,n}}{2} \left[ \left( p^{s}_{I,n} \right)^2 + \left( q^{s}_{I,n} - g^{s}_{I,n}q_{I,s} \right)^2 \right] \right\}.
\end{align}
\noindent Here we have included the system-bath couplings as  system-dependent shifts in the minima of the bath oscillators, which ensures translational invariance of the bath with respect to the system. The $g$ coefficients in the above expressions are the bilinear coupling coefficients with the form,
\begin{align}
g_{k,n} = \frac{\sqrt{2} c_{k,n}}{\hbar \omega_{k,n}},    
\end{align}
which comprise the spectral density function,
\begin{align}
J_{m,k} (\omega) = \frac{\pi}{2} \sum_n \frac{c_{k,n}^2}{\hbar \omega_{k,n}} \delta (\omega - \omega_{k,n}).
\end{align}
Here $m=\text{el,vib}$ denotes whether the spectral density corresponds to an electronically- or vibrationally-coupled environment and $k$ serves here as a composite index ($k=I$ for the electronic bath and $k=I,f/s$ for the vibrational bath) describing the environmental modes that are coupled to the different system degrees-of-freedom in Eqs. \ref{eq:elbath} and \ref{eq:vibbath}. The spectral densities are all assumed to have the Debye form,
\begin{align}
J_{m,k}(\omega) = 2 \lambda_m \gamma_m \omega \frac{1}{\gamma_m^2 + \omega^2},
\end{align}
where $\lambda_m$ is the reorganization energy and $\gamma_m$ is the bath relaxation timescale and each $m,k$ environment. These parameters are chosen such that the bath represents a weakly-coupled, Markovian bath so that the use of multilevel Redfield theory is justified in treating the dynamics of the total system-bath Hamiltonian.\cite{montoya2015extending,schile2019simulating} We note here that while this form is consistent with much of the underlying physics of the total system, it is primarily phenomenologically included to induce weak dissipation and dephasing for ease of numerical simulations and a further study that considers the effects a more systematically imposed system-bath coupling is warranted. A detailed list of the model parameters used in this study can be found in Table \ref{tab:params}.

\begin{table}[]
    \centering
    \begin{tabular}{l l}
         \hline Parameter & Value (cm$^{-1}$)  \\
         \hline \hline $\hbar \omega_{A,g,f}$ & 1650 \\
         $\hbar \omega_{B,g,f}$ & 1660 \\ 
         $\hbar \omega_{A,e,f}$ & 1545 \\ 
         $\hbar \omega_{B,e,f}$ & 1540 \\ 
         $\hbar \omega_{A,s}$ & 200 \\
         $\Delta E$ & 100 \\
         $J_0$ & 100 \\
         $\lambda_{el}$ & 35 \\
         $\lambda_{vib}$ & 17.5 \\
         $\hbar \gamma_{el}$, $\hbar \gamma_{vib}$ & $\sim$ 106 \\
         $1/\beta$ & $\sim$ 105 \\
         $S_f$ & 0.005 (dimensionless) \\
         $\mu_{A0}/\mu_{B0}$ & -4 (dimensionless)
    \end{tabular}
    \caption{Fixed parameters used in the model heterodimer. All parameters are in units of cm$^{-1}$ unless otherwise specified.}
    \label{tab:params}
\end{table}

\subsection{\label{sec:spectroscopy}Linear Absorption and 2DEV Spectroscopy from Quantum Master Equations}

To calculate spectroscopic observables we utilize the response function formalism, which has been described elsewhere\cite{mukamel1999principles}, so we restrict our discussion to the key aspects of our simulation. In this formalism, linear and nonlinear spectra can be related via Fourier Transforms of correlation functions. Specifically, for a linear absorption spectrum in the impulsive limit, the relevant response function is
\begin{align}
R(t) = \left( \frac{i}{\hbar} \right) \theta (t) \trace{\mu_{el} G(t) \mu_{el}^{\times} \rho_{eq}},
\end{align}
where $\mu^{\times} \cdot = [\mu, \cdot]$, $\trace{\cdot}$ is the quantum mechanical trace over the full system plus bath Hilbert space, $\theta (t)$ is the Heaviside step function, and $\rho_{eq}$ is the thermal equilibrium density matrix given by
\begin{align}
\rho_{eq} = \frac{e^{-\beta H}}{\trace{e^{-\beta H}}},
\end{align}
where $\beta$ is inverse thermal energy. This response function is a dipole-dipole autocorrelation function of the electronic dipole given by 
\begin{align}
\mu_{el} = \mu_A + \mu_B
\end{align}
where 
\begin{align}
\mu_I = \mu_{I0} \left( 1 + \delta_{IA} \sqrt{2} \eta q_{A,s} \right) \left( \dbraket{I}{G} + \dbraket{G}{I} \right).
\end{align}
The time-dependence is given by action of the propagator $G(t) \cdot = e^{-i H t / \hbar} \cdot e^{i H t / \hbar}$, which is the unitary evolution in the full Hilbert space. This unitary evolution is prohibitively expensive, so we utilize the quantum master equation (QME) technique whereby we take a partial trace over the bath degrees-of-freedom and compute the response function from the dynamics of the reduced density matrix\cite{fetherolf2017linear},
\begin{align}
\traceB{G(t) \mu_{el}^{\times} \rho_{eq}} = \mathcal{G}(t) \rho_{\mu}
\end{align}
where $\rho_{\mu}$ is the reduced density matrix of the system after action of the dipole operator and $\mathcal{G} (t)$ is the reduced propagator defined by our QME. The Redfield theory approach taken here uses a double perturbation theory in both the light-matter interaction and system-bath interaction, where the light-matter interaction is assumed to be even weaker than the weak system-bath coupling.\cite{albert2016geometry,levy2020response} In this representation the response function is,
\begin{align}
R(t) = \left( \frac{i}{\hbar} \right) \theta (t) \traceS{\mu_{el} \mathcal{G}(t) \rho_{\mu}}.
\end{align}
Here we also invoke the rotating wave approximation (RWA), which reduces the terms allowed in the expansion of the commutators. Denoting the dipole operators as a sum of raising and lowering dipole operators, respectively,
\begin{align}
\mu_{el}^{+} &= \mu_A^+ + \mu_B^+ \nonumber \\
&= \mu_{A0} \left( 1 + \sqrt{2} \eta q_{A,s} \right) \dbraket{A}{G} + \mu_{B0} \dbraket{B}{G} \\
\mu_{el}^- &= \left( \mu_{el}^+ \right)^{\dagger},
\end{align}
and ignoring the negative frequency contribution, the response function then becomes
\begin{align}
R(t) = \left( \frac{i}{\hbar} \right) \theta (t) \traceS{\mu_{el}^- \mathcal{G} (t) \rho_{\mu^+}},
\end{align}
where $\mathcal{G} (t) \rho_{\mu^+} = \traceB{G(t) \mu_{el}^+ \rho_{eq}}$. The corresponding linear absorption spectrum is given by the imaginary part of the Fourier transform
\begin{align}
S (\omega_{\text{exc.}}) = \text{Im} \int d t \; e^{i\omega_{\text{exc.}} t} R(t),
\end{align}
where $\omega_{\text{exc.}}$ is the excitation frequency less the excitation energy $\epsilon$.

2DEV spectroscopy is a cross-peak specific multidimensional spectroscopic technique where the signal arises from both visible and subsequent infrared light-matter interactions. Specifically, visible excitation pulses prepare an ensemble of electronic/vibronic states which evolve as a function of waiting time, $T$. The evolution of the ensemble is then tracked via an infrared detection pulse. 

Within the same formalism, the response function for 2DEV spectroscopy can be written as
\begin{align}
R_3 &(t_{\text{det.}}, T, t_{\text{exc.}}) = \left( \frac{i}{\hbar} \right)^3 \theta (t_{\text{det.}}) \theta (T) \theta (t_{\text{exc.}}) \nonumber \\
&\times \text{Tr} \left\{\mu_{vib} G(t_{\text{det.}}) \mu_{vib}^{\times} G(T) \mu_{el}^{\times} G(t_{\text{exc.}}) \mu_{el}^{\times} \rho_{eq} \right\},
\end{align}
where $t_{\text{exc.}}$ denotes the time between the two visible pulses, $t_{\text{det.}}$ denotes the time between the infrared pulses, and the vibrational dipole operator acting on the high-frequency modes is given by
\begin{align}
\mu_{vib} = \mu_{A,f} + \mu_{B,f}
\end{align}
where $\mu_{I,f} = \sqrt{2} q_{I,f} \braket{I}$ and we have ignored the vibrational TDM of the slow mode due to non-resonance with the infrared probe. We again utilize the QME technique to compute the response function, which in the weak-coupling ($\lambda_m \rightarrow 0$) and Markovian ($\gamma_m \rightarrow 0$) limits we have chosen here reduces to the expression obtained from the quantum regression theorem\cite{fetherolf2017linear,alonso2005multiple},
\begin{align}
R_3 &(t_{\text{det.}}, T, t_{\text{exc.}}) = \left( \frac{i}{\hbar} \right)^3 \theta (t_{\text{det.}}) \theta (T) \theta (t_{\text{exc.}}) \nonumber \\
&\times \traceS{ \mu_{vib} \mathcal{G} (t_{\text{det.}}) \mu_{vib}^{\times} \mathcal{G}(T) \mu_{el}^{\times} \mathcal{G}(t_{\text{exc.}}) \rho_{\mu} }.
\end{align}
Working also with the RWA invokes further simplifications, specifically to the number of pathways\cite{oliver2014correlating}, giving the response function as a sum of rephasing (RP) and non-rephasing (NR) pathways
\begin{align}
R_3 (t_{\text{det.}}, T, t_{\text{exc.}}) &= R_3^{\text{RP}} (t_{\text{det.}}, T, t_{\text{exc.}}) \nonumber \\
&+ R_3^{\text{NR}} (t_{\text{det.}}, T, t_{\text{exc.}}),
\end{align}
where, denoting $K=\text{NR},\text{RP}$,
\begin{align}
R_3^{K} (t_{\text{det.}}, T, t_{\text{exc.}}) &= R^{K}_{\text{GSB}} (t_{\text{det.}}, T, t_{\text{exc.}}) \nonumber \\
&- R^{K}_{\text{ESA}} (t_{\text{det.}}, T, t_{\text{exc.}})
\end{align}
where GSB denotes the ground-state bleach pathways given by
\begin{align}
R^{\text{RP}}_{\text{GSB}} &(t_{\text{det.}}, T, t_{\text{exc.}}) = \left( \frac{i}{\hbar} \right)^3 \theta (t_{\text{det.}}) \theta (T) \theta (t_{\text{exc.}}) \nonumber \\
&\times \traceS{ \mu_{vib}^- \mathcal{G} (t_{\text{det.}}) \mu_{vib}^+ \mathcal{G}(T) \mu_{el}^- \mathcal{G}(t_{\text{exc.}}) \rho_{\mu^+} } \\
R^{\text{NR}}_{\text{GSB}} &(t_{\text{det.}}, T, t_{\text{exc.}}) = \left( \frac{i}{\hbar} \right)^3 \theta (t_{\text{det.}}) \theta (T) \theta (t_{\text{exc.}}) \nonumber \\
&\times \traceS{ \mu_{vib}^- \mathcal{G} (t_{\text{det.}}) \mu_{vib}^+ \mathcal{G}(T) \left( \mathcal{G}(t_{\text{exc.}}) \rho^{\dagger}_{\mu^+}\right) \mu_{el}^+ } 
\end{align}
and ESA denotes the excited state absorption pathways given by
\begin{align}
R^{\text{RP}}_{\text{ESA}} &(t_{\text{det.}}, T, t_{\text{exc.}}) = \left( \frac{i}{\hbar} \right)^3 \theta (t_{\text{det.}}) \theta (T) \theta (t_{\text{exc.}}) \nonumber \\
&\times \traceS{ \mu_{vib}^- \mathcal{G} (t_{\text{det.}}) \mu_{vib}^+ \mathcal{G}(T) \left( \mathcal{G}(t_{\text{exc.}}) \rho_{\mu^+} \right) \mu_{el}^-  } \\
R^{\text{NR}}_{\text{ESA}} &(t_{\text{det.}}, T, t_{\text{exc.}}) = \left( \frac{i}{\hbar} \right)^3 \theta (t_{\text{det.}}) \theta (T) \theta (t_{\text{exc.}}) \nonumber \\
&\times \traceS{ \mu_{vib}^- \mathcal{G} (t_{\text{det.}}) \mu_{vib}^+ \mathcal{G}(T) \mu_{el}^+  \mathcal{G}(t_{\text{exc.}}) \rho^{\dagger}_{\mu^+}  }.
\end{align}
Here we have also used the raising and lowering operator representation of the vibrational dipole operator
\begin{align}
\mu_{vib}^+ &= \mu_{A,f} + \mu_{B,f} = a^{\dagger}_{A,f} \braket{A} + a^{\dagger}_{B,f} \braket{B}, \\
\mu_{vib}^- &= \left( \mu_{vib}^+\right)^{\dagger}
\end{align}
where $a^{\dagger}_{I,f}$ denotes the bosonic creation operator of the fast mode of chromophore $I$. The signal observed experimentally is then the double Fourier transform over the excitation and detection times,
\begin{align}
\chi (\omega_{\text{det.}}, T, \omega_{\text{exc.}}) = \text{Re} &\left\{ \chi^{\text{RP}} (\omega_{\text{det.}}, T, \omega_{\text{exc.}}) \right. \nonumber \\ 
&+ \left. \chi^{\text{NR}} (-\omega_{\text{det.}}, T, \omega_{\text{exc.}}) \right\},
\end{align}
where,
\begin{align}
\chi^{\text{K}} (\omega_{\text{det.}}, T, \omega_{\text{exc.}}) = \int d t_{\text{det.}} &\int d t_{\text{exc.}} e^{i(\omega_{\text{det.}}t_{\text{det.}} +  \omega_{\text{exc.}} t_{\text{exc.}}) }  \nonumber \\
&\times R^{I}_3 (t_{\text{det.}}, T, t_{\text{exc.}}).
\end{align}
The visualization of the data is typically best presented in the form of excitation frequency ($\omega_\mathrm{exc.}$)-detection frequency ($\omega_\mathrm{det.}$) correlation plots of the total absorptive spectrum parameterized by $T$. 

\subsection{\label{sec:eigs}Eigenstate Structure of the Model Hamiltonian}

The effects from the distinct vibronic coupling mecahnisms are displayed in the eigenenergy levels shown in Fig. \ref{fig:fig1}b for which we will first focus on the electronic/vibronic manifold. In the case where there is no vibronic coupling ($S=0$, $\eta=0$) we see that the lowest energy eigenstates in the excited state manifold consist of two electronically mixed states with respect to the chromophore sites denoted by a square and circle. We note that, throughout this paper, we will colloquially refer to excitonic states of particular electronic or vibronic mixing character in accordance with their assigned shapes in Fig. \ref{fig:fig1}b. There is an additional state, denoted by a star, which is similar in its site character to the lowest-energy (square) eigenstate, but has a single quantum from the low-frequency mode on the $A$ chromophore. This state is nearly degenerate with the higher-energy (circle) eigenstate, but is composed of sites that are virtually uncoupled to the aformentioned eigenstates due to the orthogonality of the vibrational states on different electronically excited states without any vibronic coupling.

When vibronic mixing is instigated through FC activity ($S=0.1$, $\eta=0$), the nearly degenerate energy eigenstates are strongly coupled and energetically split into the star state, which is a vibronically mixed state due to the additional character of multiple low-frequency vibrational states from a \textit{single} electronically excited state, and the circle state, which is still primarily electronically coupled, but has additional character of multiple low-frequency vibrational states from \textit{both} electronically excited states. We thus refer to the energy eigenstates denoted by a square and circle as electronically coupled states, while the state denoted by a star is referred to as a vibronically coupled state.

Although difficult to capture in the energy level diagram, the energetic splitting between the circle and star states increases in the HT active case ($S$=0.1, $\eta=-0.15$) versus FC active ($S$=0.1, $\eta=0$). In either scenario, the vibronic coupling clearly serves to distribute site $A$ character throughout the excited state manifold, therefore promoting additional possible relaxation pathways. HT activity, though, specifically results in the distribution of pure electronic character from site $A$ to the vibronically coupled state (star) in contrast to FC activity which only distributes vibrational (low-frequency mode) character from site $A$. In this way, in the presence of HT activity, the circle state is nearly invariant, retaining its electronic-coupling character, but the star state gains pure electronic-coupling character, unlike in the FC active scenario. While not shown in Fig. \ref{fig:fig1}b, the next set of excitonic states in the electronic/vibronic manifold are electronic replicas of the star and circle states with an additional quantum in each vibrational state of the slow mode. These unpictured states thus contribute to the intensity borrowing effect of HT activity in the absorption lineshape.

Currently, the discussion has been restricted to the electronic/vibronic manifold, however, a comparison of the site character of the excitonic states in the vibrational manifold reveals striking differences. In fact, the high-frequency excited state vibrational modes are clearly influenced by changes in relative site contributions, which makes them sensitive reporters of vibronic mixing mechanisms. The eigenstates in the vibrational manifold are also labelled by shapes denoting the predominant transitions from the electronic/vibronic manifold due to the vibrational transition dipole moment. In this manner, we note that excitonic states in the vibrational manifold with the same shape as those in the electronic/vibronic manifold have the same electronic/vibronic character. When $S$ is nonzero, transitions between these manifolds can change the electronic/vibronic character due to changes in the vibrational transition dipole moment matrix elements. A focused discussion on the interpretation of vibronic coupling through a spectroscopic interrogation of the electronic/vibronic manifold versus both the electronic/vibronic and vibrational manifolds is reserved for Sec. \ref{sec:static}.

\section{\label{sec:static}Static Signatures of Vibronic Coupling}

\begin{figure}
\includegraphics[scale=0.49]{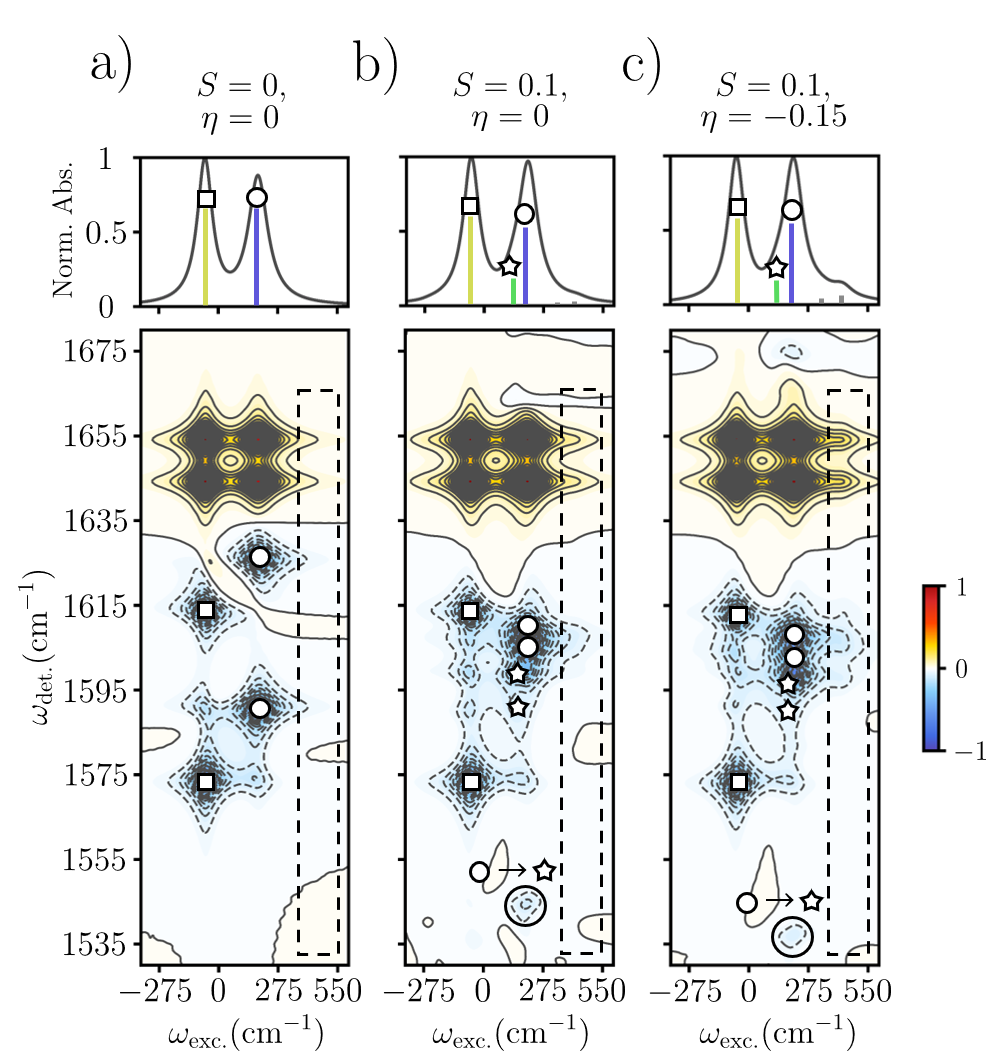}
\caption{\label{fig:fig2} (Top row) Electronic linear absorption spectra for the three treatments of vibronic coupling---a) no coupling, b) FC activity, and c) HT activity. Stick spectra are also shown where yellow (square), green (star), and blue (circle) indicate the three lowest-energy excitonic transitions, explicitly described in Fig. \ref{fig:fig1}, while gray sticks indicate higher-lying vibronic transitions. (Bottom row) Corresponding 2DEV spectra at $T=0$ fs. Positive, red/yellow features indicate GSBs and negative, blue features indicate ESAs. Contour levels are drawn in 2\% intervals. All spectra have been normalized to the maximum in each data set. ESA peaks are labeled by shapes according to transitions to the electronic/vibronic manifold as indicated in Fig. \ref{fig:fig1}. The black, dashed box highlights the higher-excitation frequency portion of the spectra where vibronic transitions appear. In b) and c), the circled ESA transition at the bottom is assigned to a transition between states of different excitonic character through a vibrational pulse.}
\end{figure}

While 2DEV spectroscopy gives a time-dependent spectroscopic signal from which dynamical phenomena can be inferred, it is first useful to uncover the ways in which it can be utilized to unravel the detailed structure arising from the underlying system Hamiltonian. In particular, we compare the signal observed from electronic linear absorption spectroscopy and the signal observed from 2DEV spectroscopy at a waiting time of $T=0$ fs. To show the specific effects arising from FC activity and HT activity we have computed both spectra with pair values of $S$ and $\eta$ at $(S,\eta) = (0,0), (0.1, 0), (0.1,-0.15)$, which are shown in Fig. \ref{fig:fig2}. When both parameters are set to zero, that is, there is neither FC nor HT activity, we expect to see coupling between the two chromophores that is purely electronic in nature. Indeed the linear absorption spectra (Fig. \ref{fig:fig2}a) shows two peaks that are inhomogeneously broadened with respect to the stick spectra due to the weak coupling between the system and bath. These peaks are transitions to the two lowest excitonic states in the excited state manifold, with zero vibrational quanta in the low-frequency modes, which have an excitonic energy gap of $\hbar \Omega_R$. The 2DEV spectrum gives additional structural information in both the GSB (positive) or ESA (negative) signals from the quartet structure owing to the correlation of the excitonic states with the vibrational character of the fast modes for each chromophore in each electronic state populated. The two excitonic transitions are observable as bands along the excitation axis with splitting equal to $\hbar \Omega_R$, however, additional cross-correlation between these bands at various positions along the detection axis is observed (see Fig. \ref{fig:fig2}a in the region spanning 1570$\sim$1595 cm$^{-1}$) which shows that the excitonic states are comprised of sites that are electronically coupled. 
The peaks along each band report on the population of particular excitonic states in electronic/vibronic manifold.
Since the high-frequency modes are local to each site, there are two vibrational peaks of the same electronic/vibronic character per band (denoted by the same shapes) that appear through coupling with excitonic states in the vibrational manifold (see Fig. \ref{fig:fig1}b).
This locality also provides some information about the relative populations in each site rather than purely excitonic populations, despite working in the electronically coherent regime. The 2DEV signal, even in this very simple case, goes well beyond the observable description obtainable by linear absorption---particularly because both the electronic/vibronic and vibrational manifolds are directly interrogated spectroscopically in the former. In this way, it is understandable how vibronic mixing mechanisms could be heavily obscured---even in other multidimensional spectroscopies---that are limited only to interrogations of the electronic/vibronic manifold. 

The stark contrast in detectable information between these spectroscopies arises in the presence of vibronic coupling activity. The linear absorption and 2DEV spectra for the FC active case ($S$=0.1, $\eta$=0) are shown in Fig. \ref{fig:fig2}b. Despite a significant change in the structure of the excitonic states, the linear absorption spectrum is virtually indistinguishable from the vibronically inactive spectrum when accounting for broadening. As is shown in the stick spectrum, the new vibronic excitonic state (star) is excited, however, due to the relative weakness of the transition and the comparable excitonic gap between the vibronic and the higher-energy electronic excitonic states (star and circle, respectively) this state is masked under typical broadening. This excitonic state is, however, clearly shown in the 2DEV spectrum. As was expected from analysis of the excitonic states (see Sec. \ref{sec:theory}), the lowest-lying excitonic state remains largely unchanged in its excitation energy and vibrational structure, however, additional structure in the cross-coupling along the detection axis of this band is observed since this excitonic state now has site character that couples to the vibronic (star) state in addition to the higher-energy electronic (circle) state. In essence, detection via the vibrational manifold serves to disperse the spectroscopic signatures of the excitonic states along the detection axis where even slight changes due to various couplings can be readily observed.

The higher-energy excitation band retains this substructure from the additional vibronic excitonic state, however, it is notable that there is a small, but detectable, energy shift along this band corresponding to the different excitonic states---the vibronic (star) state is slightly lower in energy than the electronic (circle) state. 
An additional subtle feature arises along the higher-energy excitation band at a lower detection frequency. This feature is a unique consequence of FC activity and is a signature of the site mixing in both the vibronic/electronic (star/circle, respectively) states and newly allowed transitions in the vibrational TDM. Specifically, as a result of the mixing, vibrational transitions with lower energy difference (electronic circle to vibronic star transitions in Fig. \ref{fig:fig2}b) can emerge---a transition that is expressly disallowed without FC activity due to the orthogonality of the excitonic states with respect to the low-frequency vibrational states. We also note that additional broadening in the higher-energy band is exhibited in both the GSB and ESA signals, which we attribute to coupling between the higher-energy (circle) excitonic state and other vibronic states, however, this effect is likely not distinguishable in practice.

In the final case, $(S=0.1,\eta=-0.15)$, we consider the simultaneous effect of both FC activity and HT activity on the structure of the spectra. While the vibronic state is still masked by broadening in the linear absorption spectrum, a new peak appears at an excitation energy nearly $\hbar \omega_{A,s}$ larger than the higher-energy excitonic (circle) state, which is due to the intensity borrowing effect of HT activity, i.e. there are even stronger dipole-allowed transitions to higher-lying excitonic states with additional vibrational quanta in the low-frequency mode. 
These additional transitions specifically build on the vibrational progression of the low-frequency mode in the circle and star states---rather than the square state---due to the near-resonance condition of the circle and star states in the FC inactive case.
The 2DEV spectra expectedly picks up this feature along the excitation axis in both the GSB and ESA signals, however, it is interestingly correlated with IR transitions similar to the circle state rather than the star state or a combination of the circle and star state. This correlation is due to the relative intensities that can be borrowed from the circle and star states, that is, the HT activity induces transitions that are like the circle state plus one vibrational quantum in the low-frequency mode with a stronger signal than the star state. This correlation also indicates that 2DEV spectroscopy directly reports on HT activity if the side-bands exactly replicate, with lower intensity, the lower-energy excitonic states along the detection axis and if no additional IR transitions emerge at lower detection energies akin to the circle to star IR transition from FC activity described above. 

A final point regarding the HT activity is that the observed signal here---the intensity borrowing from the dominant excitonic states along the excitation axis---is strictly due to the form of the non-Condon activity we have chosen, namely that the low-frequency mode changes the magnitude of the dipole moment and thus changes the electronic TDM directly. The same effect in the electronic coupling could arise, to first-order, from different modes that modulate the relative positions of the chromophores, but leave invariant the TDM. Since the structure of the excitonic states is apparently not influenced as much by HT activity as FC activity in the electronically coherent regime, this HT activity distinctly shows up as stronger side-band transitions along the excitation axis, which would not be present in other forms of mode-dependent electronic coupling terms.

\section{\label{sec:dynamics}Dynamical Signatures of Vibronic Coupling}

\begin{figure*}
\includegraphics[scale=0.52]{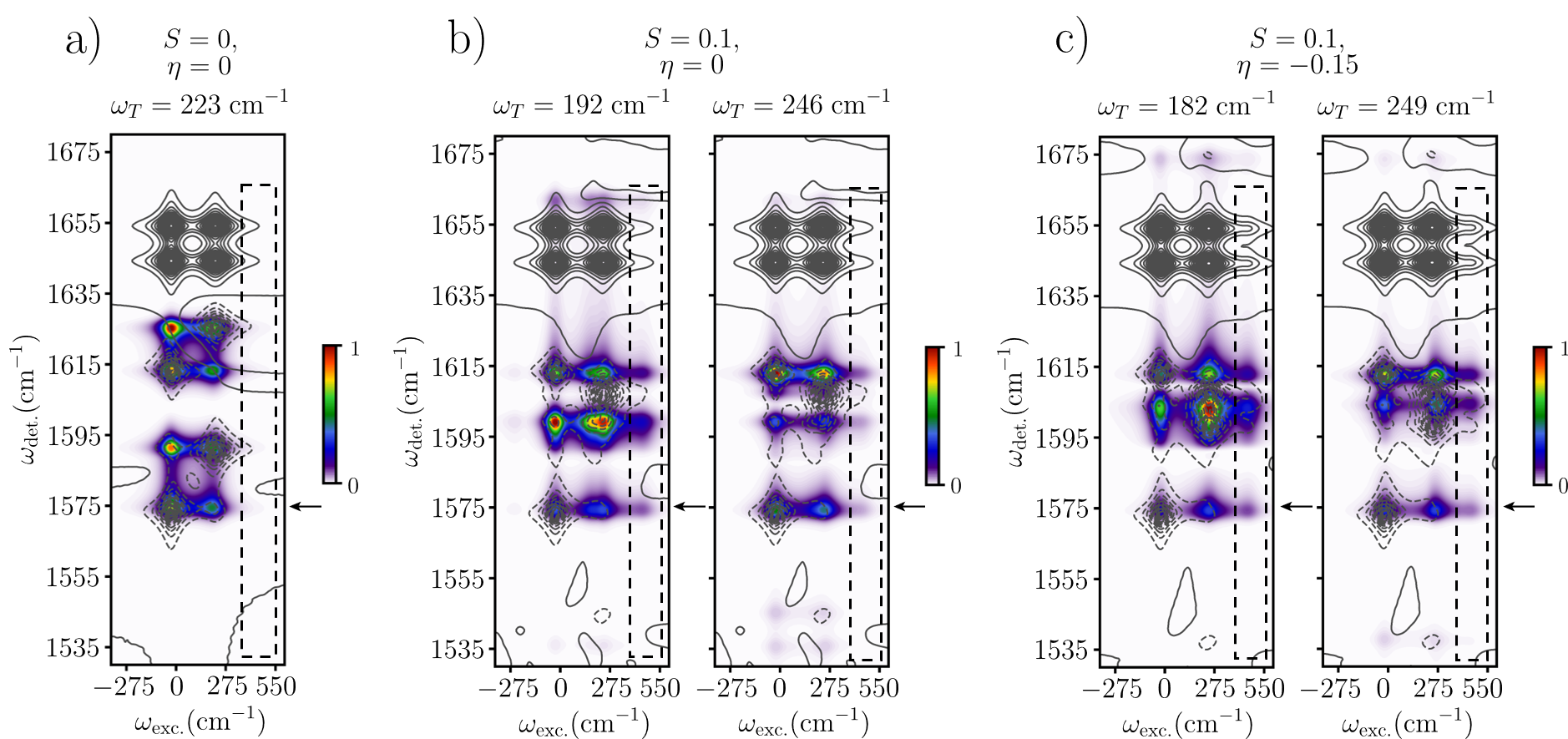}
\caption{\label{fig:fig3} Beat maps at specific $\omega_T$ values corresponding to the excitonic energy gaps in the models where there is a) no vibronic coupling, b) FC activity, and c) HT activity. For each model, the plots are normalized to the maximum beat frequency amplitude. The colormap indicates spectral regions that oscillate at the given $\omega_T$ values with amplitudes ranging from zero (white) to one (red), the maximum value. Contour lines indicate the 2DEV spectra for each model at $T=0$ fs. The black, dashed box highlights the higher-excitation frequency portion of the spectra where vibronic transitions appear. The black arrows indicate the spectral region of $\omega_\mathrm{det.}$ that is further analyzed in Fig. \ref{fig:fig5}.}
\end{figure*}

\begin{figure} 
\includegraphics[scale=0.43]{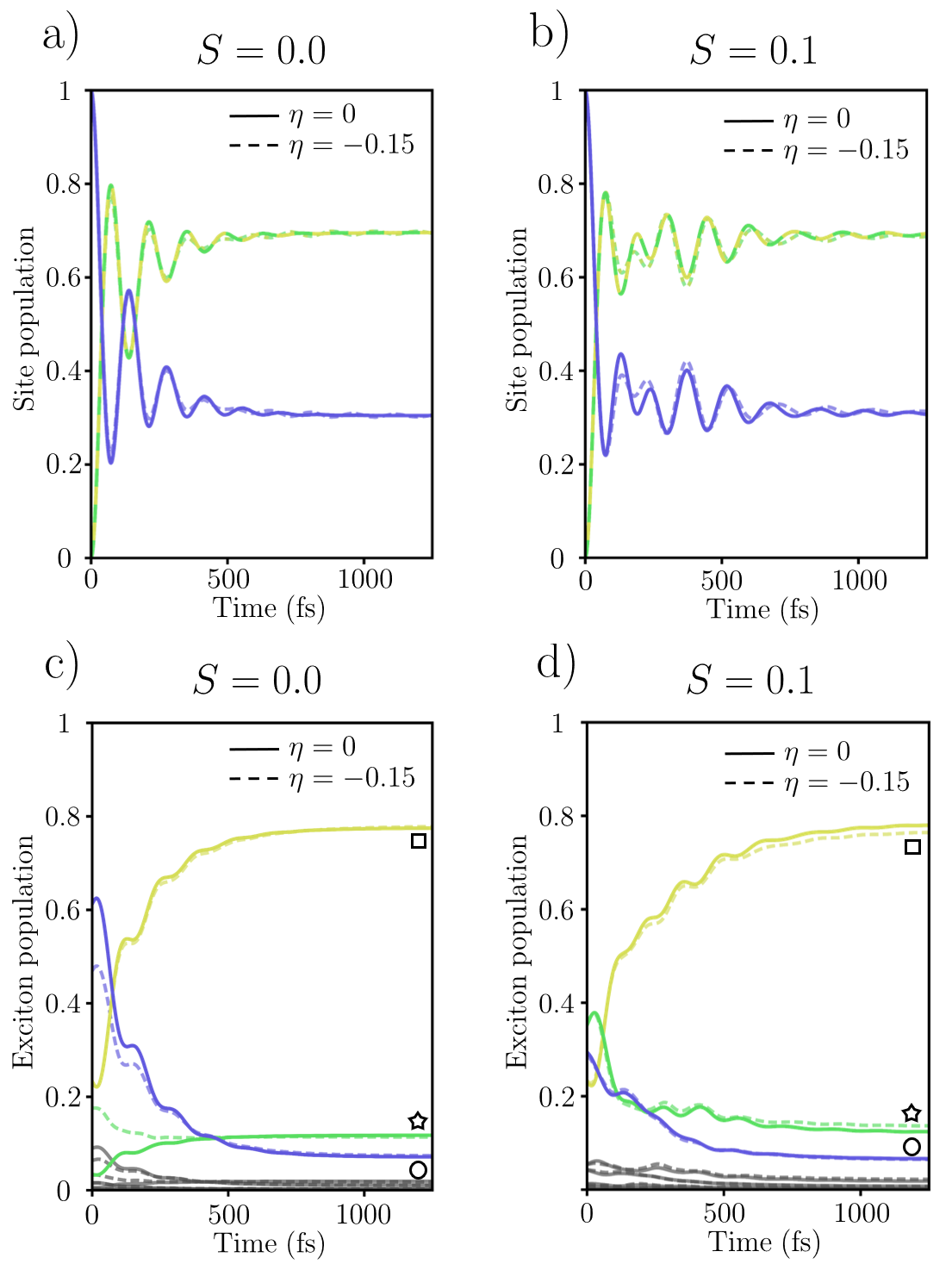}
\caption{\label{fig:fig4} (Top row) Site populations for an initially, vertically excited wavepacket into the $B$ site  where a) $S=0$ and b) $S=0.1$. Yellow/green indicates the population of site $A$ and blue indicates the population of site $B$. (Bottom row) Corresponding exciton populations where a) $S=0$ and b) $S=0.1$. Yellow (square), green (star), and blue (circle) indicate the populations of the three corresponding lowest-energy exciton levels, explicitly described in Fig. \ref{fig:fig1}, while gray indicates the populations of all higher-lying levels.  Throughout, solid lines indicate $\eta=0$ and dashed lines indicate $\eta=-0.15$ (i.e. no HT activity versus HT activity).}
\end{figure}

\begin{figure*}
\includegraphics[scale=0.45]{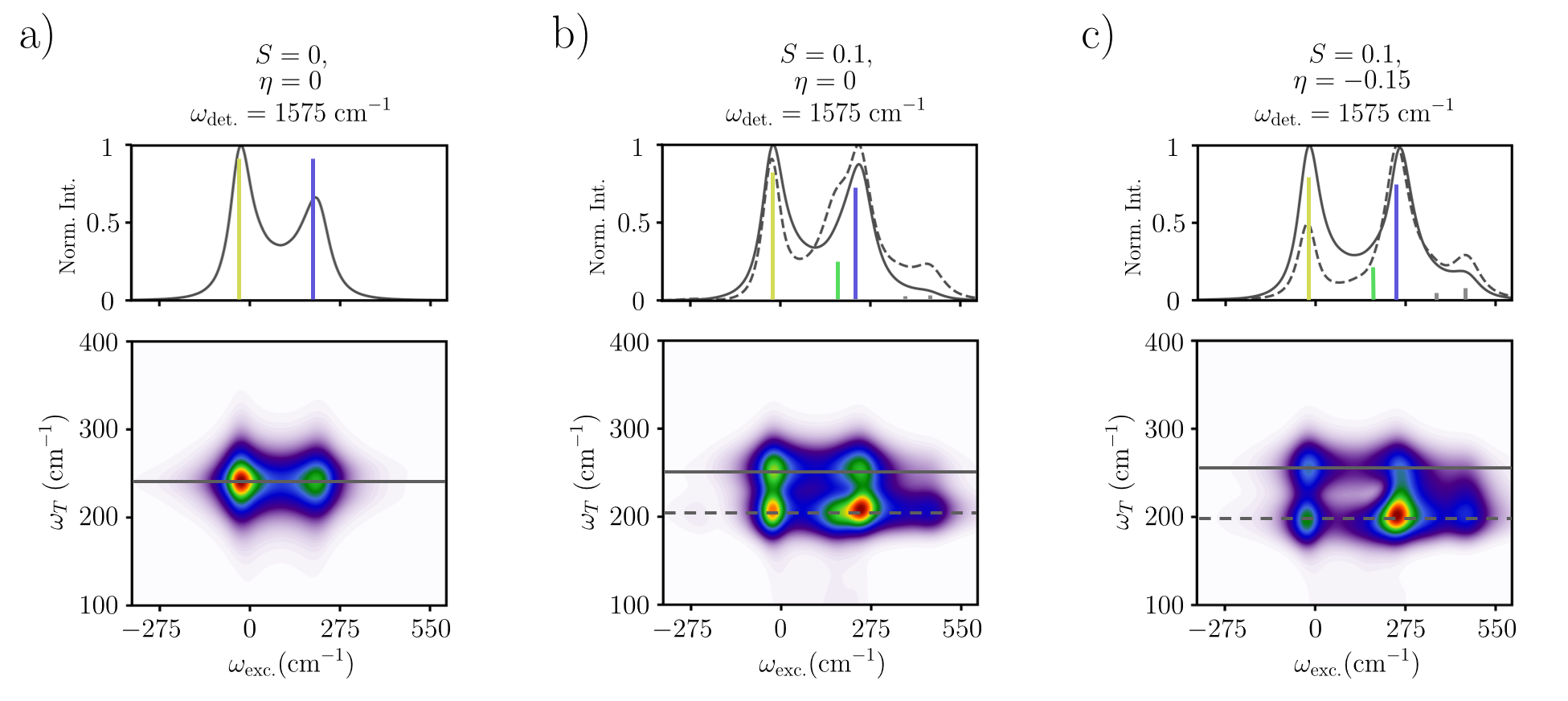}
\caption{\label{fig:fig5} Beat maps at a fixed detection frequency, $\omega_\mathrm{det.}$ (indicated by the black arrows in Fig. \ref{fig:fig3}), for the three models where there is a) no vibronic coupling, b) FC activity, and c) HT activity. The corresponding colormaps are identical to those in Fig. \ref{fig:fig3}. Slices along the excitation axis at specific beat frequencies, corresponding to the exciton energy gaps in the model, are shown above each beat map. Also shown in these plots for comparison are the electronic linear absorption stick spectra as described in Fig. \ref{fig:fig2}.}
\end{figure*}

With 2DEV spectroscopy established as a sensitive tool for witnessing vibronic effects, we turn to an analysis of how these effects manifest in the dynamics from the spectra. Rather than analyzing the complex dynamical signatures from the spectra in the time-domain, we convert to the frequency-domain to construct beat maps in the waiting time as a function of the excitation and detection frequencies. 
Specifically, these beat maps are formed by first filtering out the high-frequency oscillatory dynamics using a Savitzky-Golay filter\cite{savitzky1964smoothing}, which produces a dynamical map of the excitonic population dynamics. These population dynamics are then subtracted from the total spectra yielding the remaining coherent dynamical components (denoted by $\tilde{\chi}$) from which the power spectrum is calculated as
\begin{align}
S(\omega_{\text{det.}}, \omega_T, \omega_{\text{exc.}}) = \lvert\Xi (\omega_{\text{det.}}, \omega_T, \omega_{\text{exc.}})\rvert^{2}
\end{align}
where
\begin{align}
\Xi (\omega_{\text{det.}}, \omega_T, \omega_{\text{exc.}}) = \int dT e^{-i\omega_T T} \tilde{\chi}(\omega_{\text{det.}}, T, \omega_{\text{exc.}}).
\end{align}
In the following, we will show how these oscillatory components report directly on the interplay between excitonic states. Additionally, the conclusions drawn from this specific type of beat map analysis can be readily applied to more complex systems where the excitonic manifold as well as the dynamics are often highly congested.

Population dynamics of the sites can also be inferred from these dynamical beat maps since 2DEV probes local intramolecular modes.\cite{lewis2015method} To illustrate this point, we compare the dynamical beat maps to the population dynamics starting from an initial vertical excitation to the $B$ site given by,
\begin{align}
\rho (0) = \mu_B \rho_{eq} \mu_B^{\dagger}.
\end{align}
This initial condition considers specifically the rapid population transfer from the higher-energy $B$ site to the lower-energy $A$ site to show the complex dynamical features observed in this ultrafast process comparable to realistic systems such as LHCII. While this initial condition is not entirely physically realizable as the chromophores are intrinsically coupled and cannot be isolated in this way, it is useful to show how the dynamical signatures in 2DEV spectra are exhibited in more idealistic simulations for drawing connections between future atomistic simulations for which corresponding spectral simulations are beyond computational capabilities. 

In the beat maps, we observe peaks in the dynamical frequency $\omega_T$ that correlate with the excitonic states at particular $\omega_{\text{exc.}}$ and $\omega_{\text{det.}}$. The correlations between the dynamical frequency and the excitonic states specifically show the contribution from certain states to a particular dynamical signature, that is, which states beat at which frequencies. We have analyzed these beat maps in each parameter set $(S,\eta)$, which are shown in Fig. \ref{fig:fig3} as overlayed with the $T=0$ fs 2DEV spectra for clearer identification. In the case $(S=0,\eta=0)$ we observe a single dynamical frequency corresponding to the bare excitonic gap $\hbar \Omega_R$. This signature is to be expected as there is negligible contribution of FC activity from the high-frequency modes and no vibronic contribution from the low-frequency mode. Thus, the state populations oscillate, at times shorter than the onset of thermalization, in accordance with the dynamics of a two-level system. This beat map is consistent with the population dynamics, shown in Fig. \ref{fig:fig4}a and c, which show the site and excitonic populations, respectively. In particular, the site populations exhibit beating only at the excitonic gap between the chromophoric states with subsequent thermal relaxation. This same beating appears in the excitonic populations where it is convoluted with population transfer between the excitonic states. We have also computed the population dynamics considering only the HT activity, $(S=0,\eta=-0.15)$, and found that there is little to no difference in the site population dynamics. Rather, the difference is in the initial excitation condition of the excitonic populations due to the aforementioned change in the structure of the excitonic states to which we are exciting.

With the addition of both cases of vibronic coupling comes an additional dynamical frequency associated with quantum beating at the excitonic gap between the square and star state, which is distinct from pure Rabi oscillations. While the Rabi frequency is slightly modified, this beating frequency is still associated predominantly with the excitonic state of mostly electronic character (circle), while the additional frequency is associated with the vibronic state (star). This distinction is emphasized when considering the correlation between the beat frequency and the excitonic state character as shown in Fig. \ref{fig:fig3}b and c, which show the beat maps for the $S=0.1$ and $\eta=0,-0.15$ cases. In both cases, the modified Rabi frequency is slightly higher due to the additional coupling but in the FC-only active case, this frequency is specifically correlated with the circle state with a small contribution from the star state. This correlation is most notable when considering the lower detection frequency circle to star transition (1530$\sim$1540 cm$^{-1}$) which has a weak signal at the modified Rabi frequency but no signal at the new vibronic frequency. The vibronic frequency has much more participation from the vibronic (star) state than does the modified Rabi frequency. At this new frequency, there is also notably more activity at higher-lying vibronic states along the excitation axis suggesting that these higher-lying vibronic states are relaxing mainly to the star state. These excitation side-band correlations become significantly more prevalent in the HT active case $(S=0.1,\eta=-0.15)$. Noticeably, however, there is enhanced activity of these higher-lying vibronic states in \textit{both} frequency components. The main difference is that HT activity leads to borrowing of pure electronic character from the circle to the star state (see Fig. \ref{fig:fig1}b). This activity, in turn, leads to more equal contributions from both states at the new vibronic frequency and the (further) modified Rabi frequency facilitating participation of the higher-lying vibronic states across all beat frequencies. 

In both cases, $(S,\eta)=(0.1,0),(0.1,-0.15)$, the population dynamics (shown in Fig. \ref{fig:fig4}b and d) are virtually identical and we will thus consider them in unison. The site populations show a seemingly polychromatic beating pattern with initial electronic oscillations corresponding to the modified Rabi frequency crossing over to beating on the vibronic frequency. This pattern is also exhibited in the excitonic populations with an initial beat between the electronic (square and circle) states followed by correlated oscillations in the square and star states. In this instance, it appears as though population transfer between the chromophores is assisted by vibronic coupling, specifically FC activity, by protecting the transfer from back-oscillations. In particular, the crossover from purely electronic oscillations at short times (about one period of the modified Rabi frequency) to oscillations at the excitonic gap coupling the star state prohibits further population from transferring back to the $B$ site after transferring to the $A$ site. We emphasize, however, that this is only a weakly drawn conclusion with respect to energy transfer in realistic systems and requires further analysis in which we consider various regimes including the electronically incoherent regime. For example, the overall transfer between sites $A$ and $B$ in this case is largely dictated by the electronic coupling which distributes a reasonable amount of site $B$ character in the lowest excitonic state---in direct competition with the vibronically-induced distribution of site $A$ character among the higher-lying states. In the incoherent regime, the lowest excitonic state will almost completely resemble site $A$, however, vibronic mixing will still serve to distribute site $A$ character throughout the higher-lying excitonic states in the same way as for the models considered here (see Fig. \ref{fig:fig1}b). Therefore, we expect that vibronic effects will manifest more strongly in the incoherent regime where they are the dominant means for the distribution of site $A$ character---without the competing effects of electronic coupling distributing site $B$ character in the opposite, undesirable direction. The treatment of this regime in regards to 2DEV spectral simulations, though, is beyond the perturbative limit of Redfield theory used in this study. Nevertheless, vibronic coupling has a clear impact on the population dynamics that emerges in the dynamical signatures of the 2DEV spectra from these models. 

We further note that in both cases of vibronic mixing there are congested signals in the beat maps. It is thus useful to consider a particular slice of these beat maps along the detection axis associate with the lowest-lying excitonic state. Since this state is mostly unchanged by vibronic coupling, it can serve as a sensitive reporter of the changes in the dynamical beat frequencies through which the effects from vibronic mixing emerge. These excitonic-state specific beat maps are shown in Fig. \ref{fig:fig5}. Along with these two-dimensional beat maps we consider slices along the observed dynamical frequencies shown relative to the linear absorption stick spectrum. In the vibronically inactive case, we again observe a single dynamical frequency associated with the Rabi frequency to which both excitonic states contribute. This signature clearly identifies the connectivity between these states.\cite{arsenault2020vibronic} In systems with more complex excited state manifolds, i.e. with vibronic mixing, the implications of these maps are striking. For example, in the FC active case $(S=0.1,\eta=0)$ (Fig. \ref{fig:fig5}b) the additional peaks in the vibronic frequency band illustrate how energy flows within the excitonic manifold. By looking at slices along $\omega_T$ at the modified Rabi frequency, it is apparent that population primarily flows from the circle to square state. However, at $\omega_T$ specific to the vibronic frequency, there is an additional peak at the higher-lying vibronic side-band as well as at the star state. This distinction reveals how FC activity promotes a ``vibronic funnel" whereby excitation flows from the higher-lying states through the circle and star states down to the lowest excitonic state (square)---clearly demonstrating the additional relaxation channel. In the HT active case $(S=0.1,\eta=-0.15)$ (Fig. \ref{fig:fig5}c), we see a similar features along the lower $\omega_T$ frequency, however, in the higher $\omega_T$ value, there is amplified contribution from the higher-lying vibronic states as compared to $(S=0.1,\eta=0)$ (Fig. \ref{fig:fig5}b). This feature is perhaps a clearer demonstration of how HT activity results in additional mixing, i.e. additional vibronically-promoted relaxation pathways through the modified electronic coupling. 

\section{\label{sec:conclusion}Concluding Remarks}

In this work, we have introduced a minimal model for an electronically/vibronically coupled heterodimer for which two distinct mechanisms of vibronic coupling can be systematically tuned. This model adequately describes the coupling of a low-frequency nuclear mode to site-exciton states in a multichromophoric system and introduces a set of local high-frequency modes to report on the vibronic coupling in 2DEV spectroscopy. This low-frequency mode can induce vibronic coupling through Franck-Condon activity, which couples the nuclear mode to the site energies, or through Herzberg-Teller activity, which introduces nuclear dependence of the electronic coupling through the TDM of a single chromophore. 

Through the development of these heterodimer models, we have shown how different mechanisms of vibronic coupling, or lack thereof, manifest in both the composition of the resulting excitonic states as well as the 2DEV spectra through both static and dynamical contributions to the overall signal. In the absence of vibronic coupling, the system resembles that of a two-level model in which the dominant excitonic states are observable in the 2DEV spectra through excitation bands with vibrational structure of the chromophores and cross-peaks characterizing the electronic coupling. When the low-frequency mode is coupled to the electronic manifold, vibronic structure emerges due to an additional vibronically mixed state in the case of FC activity and an increased signal in the electronic side-band arising specifically from HT activity rather than mode-dependent electronic coupling. 2DEV spectroscopy also reports on the population dynamics due to the locality of the vibrational probe and can thus reveal nature of quantum beating patterns during energy transfer. Without vibronic coupling, the system beats at a single frequency associated with the electronic coupling while vibronic coupling introduces a new quantum beat frequency due to additional vibronically mixed excitonic states. These beat frequencies directly characterize the population dynamics and show the additional relaxation pathways vibronic coupling affords the energy transfer dynamics. Ultimately, the insight gained from this work provides a general framework for the interpretation of the underlying Hamiltonian of vibronically coupled systems. In fact, connections between previous experimental work and the present models, addressed elsewhere, have uncovered details about the vibronic coupling mechanisms in LHCII.\cite{arsenault2021note}

Various aspects do, however, require further investigation. For example, we have only considered here the electronically coherent regime where HT activity has little effect on the overall energy transfer, a feature which we do not expect to generically hold true across all regimes. With regard still to the nuclear dependence of the electronic coupling, our treatment is specific to that which arises from nuclear dependence of the dipole moment, however, a similar effect in the electronic coupling due to the spatial/orientational changes from short- or long-range nuclear fluctuations could be expected. 
A more systematic understanding of the effect on the energy transfer and the signature in 2DEV spectroscopy from these separate coupling mechanisms warrants further study. While generalizations to the model presented here would be required, the way in which electronic-nuclear coupling mechanistically mediates dynamics through conical intersections\cite{roy2020solvent,hart2020identification} or assists in charge transfer\cite{gaynor2019vibronic,yoneda2021electron,falke2014coherent} and singlet fission\cite{schultz2021influence} are similarly deserving of explicit theoretical treatment with respect to 2DEV spectroscopy.

\begin{acknowledgments}

E.A.A. and G.R.F. acknowledge support from the U.S. Department of Energy, Office of Science, Basic Energy Sciences, Chemical Sciences, Geosciences, and Biosciences Division. A.J.S. and D.T.L. were supported by the U.S. Department of Energy, Office of Science, Basic Energy Sciences, CPIMS Program Early Career Research Program under Award No. DE-FOA0002019.
E.A.A. is grateful for support from the National Science Foundation Graduate Research Fellowship (Grant No. DGE 1752814). 

\end{acknowledgments}

\section*{References}

%

\end{document}